\documentclass[runningheads, table]{llncs}

\usepackage{tikz}
\usepackage{multirow}
\usepackage{multicol}
\usepackage{xcolor}
\usepackage{booktabs}
\usepackage{float}
\usepackage{subcaption}
\usepackage{url}
\usepackage{microtype}
\usepackage{hyperref}

\usetikzlibrary{shapes.geometric, arrows, positioning, calc, arrows.meta}
\tikzstyle {input} = [trapezium, draw = black, trapezium left angle=70, trapezium right angle=110, text centered, fill = orange!25]
\tikzstyle {process} = [rectangle, draw = black, text centered, fill = blue!15]
\tikzstyle {startend} = [rectangle, draw = black, text centered, rounded corners, fill = green!25]
\tikzstyle{arrow} = [-{Stealth[length = 1.5mm, width = 1.5mm]}]

\begin{document}

\title{QuOTeS: Query-Oriented Technical Summarization}

\author{
    Juan Ramirez-Orta\inst{1} \thanks{Please send correspondence to \url{juan.ramirez.orta@dal.ca}} \and Eduardo Xamena\inst{2,3} \and Ana Maguitman\inst{5,6} \and Axel J. Soto\inst{5,6} \and Flavia P. Zanoto\inst{4} \and Evangelos Milios\inst{1}
}

\authorrunning{Ramirez-Orta et al.}

\institute{
    Department of Computer Science, Dalhousie University \and
    Institute of Research in Social Sciences and Humanities (ICSOH) \and
    Universidad Nacional de Salta - CONICET \and
    Escrever Ciência \and
    Institute for Computer Science and Engineering, UNS - CONICET \and 
    Department of Computer Science and Engineering, Universidad Nacional del Sur
}

\maketitle

%%%%%%%%%%%%%%%%%%%%%%%%%%%%%%%%%%%%%%%%%%%%%%%%%%%%

\begin{abstract}
    When writing an academic paper, researchers often spend considerable time reviewing and summarizing papers to extract relevant citations and data to compose the \emph{Introduction} and \emph{Related Work} sections. To address this problem, we propose \emph{QuOTeS}, an interactive system designed to retrieve sentences related to a summary of the research from a collection of potential references and hence assist in the composition of new papers. \emph{QuOTeS} integrates techniques from Query-Focused Extractive Summarization and High-Recall Information Retrieval to provide Interactive Query-Focused Summarization of scientific documents. To measure the performance of our system, we carried out a comprehensive user study where participants uploaded papers related to their research and evaluated the system in terms of its usability and the quality of the summaries it produces. The results show that \emph{QuOTeS} provides a positive user experience and consistently provides query-focused summaries that are relevant, concise, and complete.
    % We share the code of our system and the novel Query-Focused Summarization dataset collected during our experiments at \url{HIDDEN-IN-THIS-DRAFT-TO-ENABLE-BLIND-REVIEW}.
    We share the code of our system and the novel Query-Focused Summarization dataset collected during our experiments at \url{https://github.com/jarobyte91/quotes}.

\end{abstract}

% \footnote[]{Please send correspondence to \url{juan.ramirez.orta@dal.ca}}

% \maketitle

%%%%%%%%%%%%%%%%%%%%%%%%%%%%%%%%%%%%%%%%%%%%%%%%%%%%
\section{Introduction}
%%%%%%%%%%%%%%%%%%%%%%%%%%%%%%%%%%%%%%%%%%%%%%%%%%%%

When writing an academic paper, researchers often spend substantial time reviewing and summarizing papers to shape the \emph{Introduction} and \emph{Related Work} sections of their upcoming research. Given the ever-increasing number of academic publications available every year, this task has become very difficult and time-consuming, even for experienced researchers. A solution to this problem is to use Automatic Summarization systems, which take a long document or a collection of documents as input and produce a shorter text that conveys the same information.

The summaries produced by such systems are evaluated by measuring their fluency, coherence, conciseness, and completeness. To this end, Automatic Summarization systems can be divided into two categories, depending on their output. In Extractive Summarization, the purpose of the system is to highlight or extract passages present in the original text, so the summaries are usually more coherent and complete. On the other hand, in Abstractive Summarization, the system generates the summary by introducing words that are not necessarily in the original text. Hence, the summaries are usually more fluent and concise. Although there have been significant advances recently \cite{pegasus}, these complementary approaches share the same weakness: it is very hard for users to evaluate the quality of an automatic summary because it means that they have to go back to the original documents and verify that the system extracted the correct information.

Since evaluating summarization systems by hand is very difficult, several automatic metrics have been created with this purpose: BLEU \cite{bleu}, ROUGE \cite{rouge}, and METEOR \cite{meteor} all aim to measure the quality of the summary produced by the system by comparing it with a reference summary via the distribution of its word n-grams. Despite being very convenient and popular, all these automatic metrics have a significant drawback: since they only look at the differences in the distribution of words between the system's summary and the reference summary, they are not useful when the two summaries are worded differently, which is not necessarily a sign that the system is performing poorly.

Therefore, although Automatic Summarization systems display high performance when evaluated on benchmark datasets \cite{rush-etal-2015-neural}, they often cannot satisfy their users' needs, given the inherent difficulty and ambiguity of the task \cite{duc_2005}. An alternative approach to make systems more user-centric is Query-Focused Summarization \cite{duc_2005}, in which the users submit a query into the system to guide the summarization process and tailor it to their needs. Another alternative approach to this end is Interactive Summarization \cite{ineats}, in which the system produces an iteratively improved summary. Both of these approaches, and several others, take into account that the \emph{correct} summary given a document collection depends on both the users and what they are looking for.

\begin{figure}[ht]
    \centering
    \resizebox{0.8\columnwidth}{!}{
        \begin{tikzpicture}[node distance = 0.5cm]
            \node (sentences) [process] {Sentences};
            \node (embeddings) [process, below = of sentences] {Embeddings};
            \node (aux1) [below = 0.25cm of embeddings] {};
            \node (aux2) [below = 0.75cm of aux1] {};
            \node (ir) [process, left = 0.1cm of aux2, text width = 3.5cm] {Information Retrieval\\ Engine};
            % \node (aux) [below = 0.1 cm of ir] {};
            \node (classifier) [process, right = 0.1cm of aux2, text width = 3cm] {Machine Learning\\ Classifier};
            \node (query) [input, left = 1.25cm of ir] {Query};
            \node (documents) [input] at (query |- sentences) {Documents};
            
            % % \node (aux2) [left = 3cm of al] {};
            % \node (aux2) at (documents |- classifier) {};
            \node (recommendations) [input, below = 0.75 of ir] {Recommendations};
            \node (labels) [input, below = of recommendations]  {Labels};
            \node (aux3) [left = 0.75cm of classifier.south] {};
            \node (output) [startend, below = of labels] {Query-Focused Summary};

            \draw [black, dashed, rounded corners] ([yshift=-0.25cm, xshift=-0.25cm] ir.south west) rectangle ([yshift=0.25cm, xshift=0.25cm] sentences.north -| classifier.east);
            \node (approach) [above = 0.25 of sentences] {QuOTeS};

            \draw [arrow] (documents) -- (sentences); 
            \draw [arrow] (sentences) -- (embeddings); 
            \draw (embeddings) -- (aux1.center); 
            \draw [arrow] (aux1.center) -| (ir); 
            \draw [arrow] (aux1.center) -| (classifier); 
            \draw [arrow] (ir) -- (recommendations); 
            \draw [arrow] (recommendations) -- (labels); 
            % \draw [arrow] (embeddings) -| (classifier); 
            \draw [arrow] (labels) -| (classifier); 
            % \draw [thick] (classifier) -- (aux.center); 
            \draw [arrow] (aux3.center) |- (recommendations); 
            \draw [arrow] (labels) -- (output); 
            \draw [arrow] (query) -- (ir); 
            \draw [arrow] (query) |- (output); 
            % \draw [arrow] (aux.center) -- (output); 
        \end{tikzpicture}
    }
     \caption{Overview of how \emph{QuOTeS} works. First, the user inputs their documents into the system, which then extracts the text present in them. Next, the system splits the text into sentences and computes an embedding for each one of them. After that, the user inputs their query, which is a short paragraph describing their research, and the system retrieves the most relevant sentences using the traditional \emph{Vector Space Model}. The user then labels the recommendations and trains the system using techniques from High-Recall Information Retrieval to retrieve more relevant sentences until he or she is satisfied. Finally, the sentences labeled as relevant are returned to the user as the Query-Focused Summary of the collection.}
    \label{fig:overview}
\end{figure}
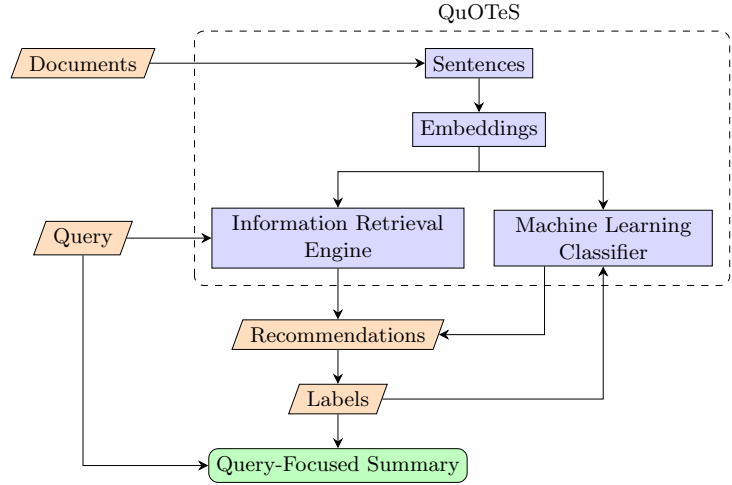

In this paper, we introduce \emph{QuOTeS}, an interactive system designed to retrieve sentences relevant to a paragraph from a collection of academic articles to assist in the composition of new papers. \emph{QuOTeS} integrates techniques from Query-Focused Extractive Summarization \cite{duc_2005} and High-Recall Information Retrieval \cite{evaluation_of_machine_learning} to provide Interactive Query-Focused Summarization of scientific documents. An overview of how \emph{QuOTeS} works and its components is shown in Fig.~\ref{fig:overview}.

The main difficulty when creating a system like \emph{QuOTeS} in a supervised manner is the lack of training data: gathering enough training examples would require having expert scientists carefully read several academic papers and manually label each one of their sentences concerning their relevance to the query, which would take substantial human effort. Therefore, we propose \emph{QuOTeS} as a self-service tool: the users supply their academic papers (usually as PDFs), and \emph{QuOTeS} provides an end-to-end service to aid them in the retrieval process. This paper includes the following contributions:

\begin{itemize}
    \item A novel Interactive Query-Focused Summarization system that receives a short paragraph (called query) and a collection of academic documents as input and returns the sentences related to the query from the documents in the collection. The system extracts the text directly from the academic documents provided by the user at runtime, minimizing the effort needed to perform complex queries on the text present in the documents. Finally, the system features techniques from High-Recall Information Retrieval to maximize the number of relevant sentences retrieved.
    \item A novel dataset composed of \emph{(Query, Document Collection)} pairs for the task of Query-Focused Summarization of Scientific Documents, each one with five documents and hundreds of sentences, along with the relevance labels produced by real users.
    \item A comprehensive analysis of the data collected during a user study of the system, where the system was evaluated using the System Usability Scale \cite{sus} and custom questionnaires to measure its usability and the quality of the summaries it produces.
\end{itemize}

%%%%%%%%%%%%%%%%%%%%%%%%%%%%%%%%%%%%%%%%%%%%%%%%%%%%
\section{Related Work}
%%%%%%%%%%%%%%%%%%%%%%%%%%%%%%%%%%%%%%%%%%%%%%%%%%%%

%%%%%%%%%%%%%%%%%%%%%%%%%%%%%%%%%%%%%%%%%%%%%%%%%%%%
\subsection{Query-Focused Summarization}
%%%%%%%%%%%%%%%%%%%%%%%%%%%%%%%%%%%%%%%%%%%%%%%%%%%%

The task of Query-Focused Summarization (QFS) was introduced in the 2005 Document Understanding Conference (DUC 2005) \cite{duc_2005}. The focus of the conference was to develop new evaluation methods that take into account the variation of summaries produced by humans. Therefore, DUC 2005 had a single, user-oriented, question-focused summarization task that allowed the community to put some time and effort into helping with the new evaluation framework. The summarization task was to synthesize a well-organized and ﬂuent answer to a complex question from a set of 25 to 50 documents. The relatively generous allowance of 250 words for each answer revealed how difﬁcult it was for the systems to produce good multi-document summaries. The two subsequent editions of the conference (DUC 2006 \cite{duc_2006} and DUC 2007 \cite{duc_2007}) further enhanced the dataset produced in the first conference and have become the reference benchmark in the field.

% Query-Chain Summarization, which combines aspects of Update Summarization and Query-Focused Summarization, was introduced in \cite{query_chain}. In this task, increasingly more specific queries about a document collection are considered, producing a new summary at every step. This process mimics an exploratory search: a user explores a new topic by submitting a sequence of queries, inspecting the summary of the resulting set, and phrasing a new query at each step. They also introduced a novel dataset comprising 22 query chains sessions of length up to 3 along their matching human summaries, each in the consumer health domain. Finally, they introduced a novel algorithm for Query-Chain Summarization based on a variation of Latent Dirichlet Allocation \cite{latent_dirichlet_allocation}. 
% The main limitation of their work is that they only focused in "zoom in" query chains, leaving "zoom out" and "switch topic" for future work. Also, they highlighted the need for a task-specific evaluation metric that exploits the structure of the chains to better assess relevance, redundancy and contrast.

Surprisingly, state-of-the-art algorithms designed for QFS do not significantly improve upon generic summarization methods when evaluated on traditional QFS datasets, as was shown in \cite{topic_concentration}. The authors hypothesized that this lack of success stems from the nature of the datasets, so they defined a novel method to quantify their Topic Concentration. Using their method, which is based on the ratio of sentences within the dataset that are already related to the query, they observed that the DUC datasets suffer from very high Topic Concentration. Therefore, they introduced TD-QFS, a new QFS dataset with controlled levels of Topic Concentration, and compared competitive baseline algorithms on it, reporting a solid improvement in performance for algorithms that model query relevance instead of generic summarizers. Finally, they presented three novel QFS algorithms (RelSum, ThresholdSum, and TFIDF-KLSum) that outperform, by a large margin, state-of-the-art QFS algorithms on the TD-QFS dataset. 
% Although their novel methods clearly outperform the previous state of the art, there is still a large gap when compared against the golden retrieval method, which the authors believe can be decreased by developing joint models that combine relevance, centrality and redundancy avoidance in a single model.

A novel, unsupervised query-focused summarization method based on random walks over the graph of sentences in a document was introduced in \cite{localized_sentence_representation}. 
First, word importance scores for each target document are computed using a word-level random walk. Next, they use a siamese neural network to optimize localized sentence representations obtained as the weighted average of word embeddings, where the word importance scores determine the weights. Finally, they conducted a sentence-level query-biased random walk to select a sentence to be used as a summary. In their experiments, they constructed a small evaluation dataset for QFS of scientiﬁc documents and showed that their method achieves competitive performance compared to other embeddings.
% Interestingly, the authors observed that in certain cases, the output summaries for different queries were the same, indicating that hyper-parameters of the method could be further adjusted. In addition, they observed cases where the query expansion failed to capture the original intention of the user.

%%%%%%%%%%%%%%%%%%%%%%%%%%%%%%%%%%%%%%%%%%%%%%%%%%%%
\subsection{High-Recall Information Retrieval}
%%%%%%%%%%%%%%%%%%%%%%%%%%%%%%%%%%%%%%%%%%%%%%%%%%%%

A novel evaluation toolkit that simulates a human reviewer in the loop was introduced in \cite{evaluation_of_machine_learning}. The work compared the eﬀectiveness of three Machine Learning protocols for Technology-Assisted Review (TAR) used in document review for legal proceedings. It also addressed a central question in the deployment of TAR: should the initial training documents be selected randomly, or should they be selected using one or more deterministic methods, such as Keyword Search? To answer this question, they measured Recall as a function of human review eﬀort on eight tasks. 
Their results showed that the best strategy to minimize the human eﬀort is to use keywords to select the initial documents in conjunction with deterministic methods to train the classifier.
% Their results showed that entirely deterministic training methods require signiﬁcantly less human review eﬀort in order to achieve any given level of Recall when compared against Passive Learning. They also found that signiﬁcantly less human review eﬀort is required when keywords are used instead of Random Sampling to select the initial documents. 
% In addition, they showed that Continuous Active Learning with Relevance Feedback displays better results than Simple Active Learning with Uncertainty Sampling, which avoids the issue of determining when training should stop. 
% The main limitation of the work is that they only studied cases where the amount of relevant documents was very low (0.25\% to 3.92\%), which is typical for the legal matters that the authors have encountered in their career. Also, they highlight that when the amount of relevant documents exceeds 10\%, the amount of labeling effort and computational requirements for the experiments would become unbearable.

Continuous Active Learning achieves high Recall for TAR, not only for an overall information need but also for various facets of that information, whether explicit or implicit, as shown in \cite{multi_faceted}. Through simulations using Cormack and Grossman’s Technology-Assisted Review Evaluation Toolkit \cite{evaluation_of_machine_learning}, the authors showed that Continuous Active Learning, applied to a multi-faceted topic, eﬃciently achieves high Recall for each facet of the topic. Their results also showed that Continuous Active Learning may achieve high overall Recall without sacrificing identiﬁable categories of relevant information. 
% Nonetheless, the authors faced the important obstacles during their research: measuring Recall is problematic because the imprecision of the definition of relevance, and because of the effort, bias and imprecision associated with sampling. This means that in this kind of scenarios, it is very important that users have a clear understanding of what makes a document relevant.  

A scalable version of the Continuous Active Learning protocol (S-CAL) was introduced in \cite{scalability}. This novel variation requires $O(log(N)) $ labeling eﬀort and $O(N log(N) )$ computational eﬀort — where $N$ is the number of unlabeled training examples — to construct a classiﬁer whose eﬀectiveness for a given labeling cost compares favorably with previously reported methods. At the same time, S-CAL oﬀers calibrated estimates of Class Prevalence, Recall, and Precision, facilitating both threshold setting and determination of the adequacy of the classiﬁer. 
% The main limitation of the work is that a number of design choices were made in the experiments presented in the paper, which were not explored further. The questions of what is a "well-defined" topic, the hyper-parameters of the algorithm, and the optimal type of sampling remain open, which the authors left for future work. 

%%%%%%%%%%%%%%%%%%%%%%%%%%%%%%%%%%%%%%%%%%%%%%%%%%%%
\subsection{Interactive Query-Focused Summarization}
%%%%%%%%%%%%%%%%%%%%%%%%%%%%%%%%%%%%%%%%%%%%%%%%%%%%

A novel system that provides summaries for Computer Science publications was introduced in \cite{erera}. Through a qualitative user study, the authors identified the most valuable scenarios for discovering, exploring, and understanding scientific documents. Based on these findings, they built a system that retrieves and summarizes scientific documents for a given information need, either in the form of a free-text query or by choosing categorized values such as scientific tasks, datasets, and more. The system processed 270,000 papers to train its summarization module, which aims to generate concise yet detailed summaries. Finally, they validated their approach with human experts. 
% Nonetheless, although their initial results are promising, the work lacks a rigorous user study to measure the effectiveness of the system.

A novel framework to incorporate users' feedback using a social robotics platform was introduced in \cite{social_robot}. Using the \emph{Nao} robot (a programmable humanoid robot) as the interacting agent, they captured the user's expressions and eye movements and used it to train their system via Reinforcement Learning. The whole approach was then evaluated in terms of its adaptability and interactivity.
% Despite the favorable results shown by the system during the experiments, the authors highlighted that the initial retrieval process can be further improved and that the quality of the summaries produced by the system still needs to be evaluated based on their usefulness to the user.

% A novel system for custom summarization of large text corpora at interactive speed was introduced in \cite{hattasch}. The system consists of a sampling component and a learned model that produces the summary. The main contribution of the work is a collection of sampling strategies and their comparison. The proposed system can provide a similar level of quality as existing summarization models that work on the entire corpus but cannot process it at interactive speed. 
% Although their results on the DUC06 and DUC07 datasets are favorable, a number of design choices of the system were left unexplored: for instance, the sampling strategy and how the system can be evaluated using abstractive summarization data remain as important future directions of work.

A novel approach that exploits the user's opinion in two stages was introduced in \cite{bayatmakou}. First, the query is refined by user-selected keywords, key phrases, and sentences extracted from the document collection. Then, it expands the query using a Genetic Algorithm, which ranks the final set of sentences using Maximal Marginal Relevance. To assess the performance of the proposed system, 45 graduate students in the field of Artificial Intelligence filled out a questionnaire after using the system on papers retrieved from the Artificial Intelligence category of The Web of Science. Finally, the quality of the final summaries was measured in terms of the user's perspective and redundancy, obtaining favorable results. 
% Nonetheless, the authors mentioned that the system can still be improved in a number of ways that have worked well in similar systems: for example, including relevance feedback in the initial retrieval stage and using different criteria to evaluate the relevance of sentences for the genetic algorithm are interesting alternatives that could further improve the performance of the system.

%%%%%%%%%%%%%%%%%%%%%%%%%%%%%%%%%%%%%%%%%%%%%%%%%%%%
\section{Design Goals}
%%%%%%%%%%%%%%%%%%%%%%%%%%%%%%%%%%%%%%%%%%%%%%%%%%%%

As shown in the previous section, there is a clear research gap in the literature: on the one hand, there exist effective systems for QFS, but on the other hand, none of them includes the user's feedback about the relevance of each sentence present in the summary. On top of that, the task of QFS of scientific documents remains a fairly unexplored discipline, given the difficulty of extracting the text present in academic documents and the human effort required to evaluate such systems, as shown by \cite{localized_sentence_representation}. Considering these limitations and the guidelines obtained from an expert consultant in scientific writing from our team, we state the following design goals behind the development of \emph{QuOTeS}:

\begin{enumerate}
    \item \textbf{Receive a paragraph query and a collection of academic documents as input and return the sentences relevant to the query from the documents in the collection}. Unlike previous works, \emph{QuOTeS} is designed as an assistant in the task of writing \emph{Introduction} and \emph{Related Work} sections of papers in the making. To this end, the query inputted into the system is a short paragraph describing the upcoming work, which is a much more complex query than the one used in previous systems. 
    
    \item \textbf{Include the user in the retrieval loop}. As shown by previous works, summarization systems benefit from being interactive. Since it is difficult to express all the information need in a single query, the system needs to have some form of adaptation to the user, either by requiring more information about the user's need (by some form of query expansion) or by incorporating the relevance labeling in the retrieval process. 
    
    \item \textbf{Provide a full end-to-end user experience in the sentence extraction process}. So far, query-focused summarization systems have been mainly evaluated on data from the DUC conferences. A usable system should be able to extract the text from various documents provided by the user, which can only be determined at runtime. Since the main form to distribute academic documents is PDF files, the system needs to be well adapted to extract the text in the different layouts in academic publications.
    
    \item \textbf{Maximize Recall in the retrieval process}. Since the purpose of the system is to help the user retrieve the (possibly very) few relevant sentences from the hundreds of sentences in the collection, Recall is the most critical metric when using a system like \emph{QuOTeS}, as users can always refine the output summary to adapt it to their needs. Therefore, we use Continuous Active Learning \cite{evaluation_of_machine_learning} as the training procedure for the classifier inside \emph{QuOTeS}.
\end{enumerate}

%%%%%%%%%%%%%%%%%%%%%%%%%%%%%%%%%%%%%%%%%%%%%%%%%%%%
\section{System Design}
%%%%%%%%%%%%%%%%%%%%%%%%%%%%%%%%%%%%%%%%%%%%%%%%%%%%

\begin{figure}[ht]
    \centering
    \resizebox{\columnwidth}{!}{
        \begin{tikzpicture}[node distance = 0.5cm]
            \node (tutorial) [process, dashed] {Tutorial};
            \node (upload) [startend, below = of tutorial] {Upload};
            \node (documents) [process, right = of upload] {Documents};
            \node (search) [process, right = of documents] {Search};
            \node (explore) [process, right = of search] {Explore};
            \node (aux) [above = of explore] {};
            \node (history) [process, right = of explore] {History};
            \node (results) [startend, right = of history] {Results};
            \draw [arrow, dashed] (tutorial) -- (upload); 
            \draw [arrow] (upload) -- (documents); 
            \draw [arrow] (documents) -- (search); 
            \draw [arrow] (search) -- (explore); 
            \draw [arrow] (explore) -- (history); 
            \draw [arrow] (history) -- (results); 
            \draw (history) |- (aux.center); 
            \draw (explore) -- (aux.center); 
            \draw [arrow] (aux.center) -| (search); 
        \end{tikzpicture}
    }
    \caption{Methodology of the system and its workflow.}
    \label{fig:tabs}
\end{figure}
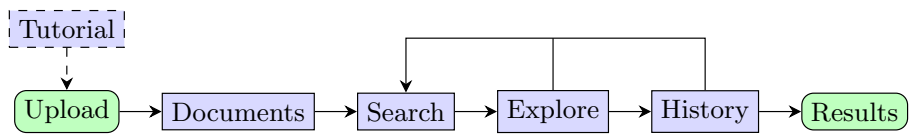

\emph{QuOTeS} is a browser-based interactive system built with \emph{Python}, mainly using the \emph{Dash} package \cite{dash}. The methodology of the system is organized into seven steps that allow the users to upload, search and explore their documents. An overview of how the steps relate to each other is shown in Fig.~\ref{fig:tabs}.

%%%%%%%%%%%%%%%%%%%%%%%%%%%%%%%%%%%%%%%%%%%%%%%%%%%%
\subsection{\emph{Tutorial}}
%%%%%%%%%%%%%%%%%%%%%%%%%%%%%%%%%%%%%%%%%%%%%%%%%%%%

In this step, the user can watch a 5-minute video\footnote[1]{The video can be watched here: \url{https://www.youtube.com/watch?v=zR9XisDFQ7w}} explaining the task that \emph{QuOTeS} was made for and an overview of how to use the system. The main part of the video explains the different parts of the system and how they are linked together. It also explains the effect of the different retrieval options and how to download the results from the system to keep analyzing them. Since users will not necessarily need to watch the video every time they use the system, the first step they see when they access the website is the \emph{Upload}, described below.

%%%%%%%%%%%%%%%%%%%%%%%%%%%%%%%%%%%%%%%%%%%%%%%%%%%%
\subsection{\emph{Upload}}
%%%%%%%%%%%%%%%%%%%%%%%%%%%%%%%%%%%%%%%%%%%%%%%%%%%%

% In this step, the users upload their documents and get the system ready to start interacting with them. A screenshot of this step is shown in Figure \ref{fig:upload}.

% \begin{figure}[ht]
%     \centering
%     \includegraphics[width=\textwidth, trim={0 300 0 0}, clip]{images/tabs/tab_upload.png}
%     \caption{ The Upload step.}
%     \label{fig:upload}
% \end{figure}

% The upload process starts with the first three components: the \emph{Upload} button, the status bar, and the \emph{Add} button. First, the users need to click on \emph{Upload}, which triggers an upload form to select the file to input into the system, which can be a single PDF, a plain text (txt) or a JSON file. 

% After selecting the file, its name is shown in the status bar so the user can verify it. Once the file is in the system, clicking on \emph{Add} extracts the text from it and adds it to the state of the system, using the \emph{Poppler} library \cite{poppler} when using PDFs and just decoding it as a UTF-8 string otherwise. Once the text from the document has been extracted, it is tokenized into sentences using the \emph{NLTK Tokenize} library \cite{nltk}. It is important to note that since the system can process documents in many languages, it needs to be aware of the proper character set of the text, which is determined automatically using the \emph{chardet} library \cite{chardet}.

In this step, the users can upload their documents and get the system ready to start interacting with them via a file upload form. Once the text from all the documents has been extracted, they can click on \emph{Process Documents} to prepare the system for the retrieval process. After that, they can select the options for the system in the \emph{Settings} screen, which contains two drop-down menus. In the \emph{Embeddings} menu, the user can choose how the system represents the query and the documents from three options: TFIDF embeddings based on word unigrams, TFIDF embeddings based on character trigrams and Sentence-BERT embeddings \cite{sentence_bert}. In the \emph{Classifier} menu, the user can choose which Supervised Machine Learning algorithm to use as the backbone for the system from three options: Logistic Regression, Random Forest, and Support Vector Machine.
% Term Frequency-Inverse Document Frequency (TFIDF) embeddings based on word unigrams are the fastest representation to compute, but they also require exact matches while searching, which makes them useful when looking for specific words or with long queries. TFIDF embeddings based on character trigrams are somewhat slower to compute than the word unigram ones, but they are also more flexible, making them a good candidate when working with a short query. Finally, the Sentence-BERT embeddings \cite{sentence_bert} are the slowest representation to compute, but they allow the system to model the relationship between single words and whole sentences, turning them into a good default option.

% In the \emph{Classifier} menu, the user can choose which Supervised Machine Learning algorithm to use as the backbone for the system from three options: Logistic Regression, Random Forest, and Support Vector Machine. 
% These options are adequate in the following use cases: when reviewing extensive collections of papers, Logistic Regression with word unigram embeddings is the best option, given its speed; in the standard scenario of reviewing between one and ten papers, Random Forest with character trigram embeddings gives a good performance and decent speed; and when reviewing one or two papers, Support Vector Machine with Sentence-BERT embeddings allow the system to model complex relationships between words and sentences, like paraphrasing and synonymy.

%%%%%%%%%%%%%%%%%%%%%%%%%%%%%%%%%%%%%%%%%%%%%%%%%%%%
\subsection{\emph{Documents}}
%%%%%%%%%%%%%%%%%%%%%%%%%%%%%%%%%%%%%%%%%%%%%%%%%%%%

In this step, the user can browse the text extracted from the documents. The sentences from the papers are shown in the order they were found so that the user can verify that the text was extracted correctly. The user can select which documents to browse from the drop-down menu at the top, which displays all the documents that have been uploaded to the system. Later on, when the user starts labeling the sentences with respect to the query, they are colored accordingly: green (for relevant) or pink (for irrelevant). 
% A screenshot of this step is shown in Figure \ref{fig:documents}.

% \begin{figure}[ht]
%     \centering
%     % \includegraphics[width=\textwidth]{images/tabs/tab_documents.png}
%     \includegraphics[width=\textwidth, trim={0 350 0 0}, clip]{images/tabs/tab_documents.png}
%     \caption{ The Documents step.}
%     \label{fig:documents}
% \end{figure}

%%%%%%%%%%%%%%%%%%%%%%%%%%%%%%%%%%%%%%%%%%%%%%%%%%%%
\subsection{\emph{Search}}
%%%%%%%%%%%%%%%%%%%%%%%%%%%%%%%%%%%%%%%%%%%%%%%%%%%%

This is the first main step of the system. In the text box, users can write their query. After clicking on \emph{Search}, the system retrieves the most relevant sentences using the classical \emph{Vector Space Model} from Information Retrieval. 
% A screenshot of this step is shown in Figure \ref{fig:search}.

The sentences below are the best matches according to the query and the representation the user picked in the \emph{Upload} step. The user can label them by clicking on them, which are colored accordingly: green (for relevant) or pink (for irrelevant). Once the users label the sentences, they can click on \emph{Submit Labels}, after which the system records them and shows a new batch of recommendations.

% \begin{figure}[ht]
%     \centering
%     \includegraphics[width=\textwidth, trim={0 325 0 0}, clip]{images/tabs/tab_search.png}
%     \caption{ The Search step.}
%     \label{fig:search}
% \end{figure}

%%%%%%%%%%%%%%%%%%%%%%%%%%%%%%%%%%%%%%%%%%%%%%%%%%%%
\subsection{\emph{Explore}}
%%%%%%%%%%%%%%%%%%%%%%%%%%%%%%%%%%%%%%%%%%%%%%%%%%%%

This is the second main step of the system. Here, the system trains its classifier using the labels the user submits to improve its understanding of the query. Two plots at the top show the distribution of the recommendation score and how it breaks down by document to help the user better understand the collection. The sentences below work exactly like in \emph{Search}, allowing the user to label them by clicking on them and submitting them into the system by clicking on \emph{Submit Labels}. Users can label the collection as much as they want, but the recommended criterion is to stop when the system has not recommended anything relevant in three consecutive turns, shown in the colored box at the top right. 
% A screenshot of this step is shown in Figure \ref{fig:explore}.

% \begin{figure}[ht]
%     \centering
%     \includegraphics[width=\textwidth, trim={0 150 0 0}, clip]{images/tabs/tab_explore.png}
%     \caption{ The Explore step.}
%     \label{fig:explore}
% \end{figure}

%%%%%%%%%%%%%%%%%%%%%%%%%%%%%%%%%%%%%%%%%%%%%%%%%%%%
\subsection{\emph{History}}
%%%%%%%%%%%%%%%%%%%%%%%%%%%%%%%%%%%%%%%%%%%%%%%%%%%%

In this step, users can review what they have labeled and where to find it in the papers. The sentences are shown in the order they were presented to the user, along with the document they came from and their sentence number to make it easier to find them. Like before, the user can click on a sentence to relabel it if necessary, which makes it change color accordingly. There are two buttons at the top: \emph{Clear} allows the user to restart the labeling process, and \emph{Download .csv} downloads the labeling history as a CSV file for further analysis. 
% A screenshot of this step is shown in Figure \ref{fig:history}.
 
% \begin{figure}[ht]
%     \centering
%     \includegraphics[width=\textwidth, trim={0 350 0 0}, clip]{images/tabs/tab_history.png}
%     \caption{ The History step.}
%     \label{fig:history}
% \end{figure}

%%%%%%%%%%%%%%%%%%%%%%%%%%%%%%%%%%%%%%%%%%%%%%%%%%%%
\subsection{\emph{Results}}
%%%%%%%%%%%%%%%%%%%%%%%%%%%%%%%%%%%%%%%%%%%%%%%%%%%%

In the last step of  \emph{QuOTeS}, the user can assess the results. There are two plots at the top that show the label counts and how they break down by document, while the bottom part displays the query and the sentences labeled as relevant. The query along these sentences make up the final output of the system, which is the Query-Focused Summary of the collection. The user can download this summary as a \emph{.txt} file or the whole state of the system as a JSON file for further analysis. 
% A screenshot of this step is shown in Figure \ref{fig:results}.

% \begin{figure}[ht]
%     \centering
%     \includegraphics[width=\textwidth, trim={0 300 0 0}, clip]{images/tabs/tab_results.png}
%     \caption{ The Results step.}
%     \label{fig:results}
% \end{figure}

%%%%%%%%%%%%%%%%%%%%%%%%%%%%%%%%%%%%%%%%%%%%%%%%%%%%
\section{Evaluation}
%%%%%%%%%%%%%%%%%%%%%%%%%%%%%%%%%%%%%%%%%%%%%%%%%%%%

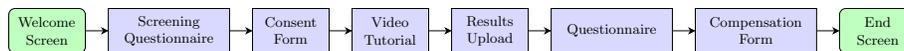
\begin{figure}[ht]
    \centering
    \resizebox{\columnwidth}{!}{
        \begin{tikzpicture}[node distance = 0.5cm, text width = 1.5cm, minimum height = 1cm]
            \node (welcome) [startend] {Welcome Screen};
            % \node (screening) [process, right = of welcome] {Screening Questionnaire};
            \node (screening) [process, right = of welcome, text width = 2.5cm] {Screening Questionnaire};
            % \node (screening) [process, right = of welcome, text width = 2.5cm] {Screening Questionnaire};
            \node (consent) [process, right = of screening] {Consent Form};
            \node (video) [process, right = of consent] {Video Tutorial};
            \node (results) [process, right = of video] {Results Upload};
            \node (questionnaire) [process, right = of results, text width = 2.5cm] {Questionnaire};
            \node (compensation) [process, right = of questionnaire, text width = 2.5cm] {Compensation Form};
            % \node (end) [startend, right = of compensation] {End Screen};
            \node (end) [startend, right = of compensation] {End Screen};
            
            \draw [arrow] (welcome) -- (screening); 
            \draw [arrow] (screening) -- (consent); 
            \draw [arrow] (consent) -- (video); 
            \draw [arrow] (video) -- (results); 
            \draw [arrow] (results) -- (questionnaire); 
            \draw [arrow] (questionnaire) -- (compensation); 
            \draw [arrow] (compensation) -- (end); 
            % \draw [arrow, dashed] (results) -| (search); 
        \end{tikzpicture}
    }
    \caption{Overview of the user study.}
    \label{fig:user_study}
\end{figure}

To evaluate the effectiveness of \emph{QuOTeS}, we performed a user study where each participant uploaded up to five documents into the system and labeled the sentences in them for a maximum of one hour. The user study was implemented as a website written using the \emph{Flask} package \cite{flask}, where the participants went through eight screens to obtain their consent, explain the task to them and fill out a questionnaire about their perception of the difficulty of the task and the performance of \emph{QuOTeS}. An overview of the user study is shown in Figure \ref{fig:user_study}.

%%%%%%%%%%%%%%%%%%%%%%%%%%%%%%%%%%%%%%%%%%%%%%%%%%%%
\subsection{Methodology}

In the \emph{Welcome Screen}, the participants were shown a quick overview of the whole user study and its duration. In the \emph{Screening Questionnaire}, they filled out a short questionnaire indicating their education level and the frequency they read academic papers. In the \emph{Consent Form} screen, they read a copy of the consent form and agreed to participate by clicking on a checkbox at the end. In the \emph{Video Tutorial} screen, they watched a five-minute video about the task and how to use \emph{QuOTeS}. In the \emph{Results Upload} screen, they were redirected to the website of \emph{QuOTeS} and after using the system for a maximum of one hour, they uploaded the JSON file containing the state of the system at the end of their interaction. In the \emph{Questionnaire} screen, they filled in a three-part questionnaire to evaluate the usability of \emph{QuOTeS}, its features and the quality of the summaries. In the \emph{Compensation Form}, they provided their name and email to be able to receive the compensation for their participation. Finally, the \emph{End Screen} indicated that the study was over and they could close their browser.

%%%%%%%%%%%%%%%%%%%%%%%%%%%%%%%%%%%%%%%%%%%%%%%%%%%%
\subsection{Participants}
%%%%%%%%%%%%%%%%%%%%%%%%%%%%%%%%%%%%%%%%%%%%%%%%%%%%

To recruit participants, we sent a general email call to our faculty, explaining the recruiting process and the compensation. To verify that participants were fit for our study, they filled out a screening questionnaire with only two questions, with the purpose of knowing their research experience and the frequency they normally read academic papers. The requirements to participate were to have completed at least an undergraduate degree in a university and to read academic papers at least once a month. The results of the screening questionnaire for the participants who completed the full study are shown in Table \ref{tab:screening_questionnaire}, while the full results of the screening questionnaire can be found in the code repository. 

\begin{table}[ht]
    \caption{Responses of the Screening Questionnaire from the participants that completed the study.}
    \centering
    \rowcolors{2}{lightgray!50}{} 
    \begin{tabular}{l|r|r}
        \toprule
        \multirow{2}{*}{Paper Reading Frequency} & \multicolumn{2}{c}{Education} \\
         &  Undergraduate &  Graduate \\
        \midrule
        Every day                     &                      1 &        4 \\
        At least once a week          &                      2 &        3 \\
        At least once every two weeks &                      0 &        1 \\
        At least once a month         &                      3 &        1 \\
        \bottomrule
    \end{tabular}
    \label{tab:screening_questionnaire}
\end{table}

%%%%%%%%%%%%%%%%%%%%%%%%%%%%%%%%%%%%%%%%%%%%%%%%%%%%
\subsection{Research Instrument}
%%%%%%%%%%%%%%%%%%%%%%%%%%%%%%%%%%%%%%%%%%%%%%%%%%%%

During the user study, the participants filled out a questionnaire composed of thirty questions divided into three parts: \emph{Usability}, \emph{Features}, and \emph{Summary Quality}. In the \emph{Usability} part, they filled out the questionnaire from the standard \emph{System Usability Scale} \cite{sus}, which is a quick and simple way to obtain a rough measure of the perceived usability of the system in the context of the task it is being used for. In the \emph{Features} part, they answered sixteen questions about how difficult the task was and the usefulness of the different components of the system. In the \emph{Summary Quality} part, they answered four questions about the relevance of the sentences in the system and the conciseness, redundancy, and completeness of the summaries produced. Finally, the participants submitted their opinions about the system and the user study in a free-text field. The full questionnaire presented to the participants can be found in the code repository.

%%%%%%%%%%%%%%%%%%%%%%%%%%%%%%%%%%%%%%%%%%%%%%%%%%%%
\subsection{Experimental Results}
%%%%%%%%%%%%%%%%%%%%%%%%%%%%%%%%%%%%%%%%%%%%%%%%%%%%

\begin{figure}[ht]
    \centering
    \includegraphics[width = 0.9\textwidth]{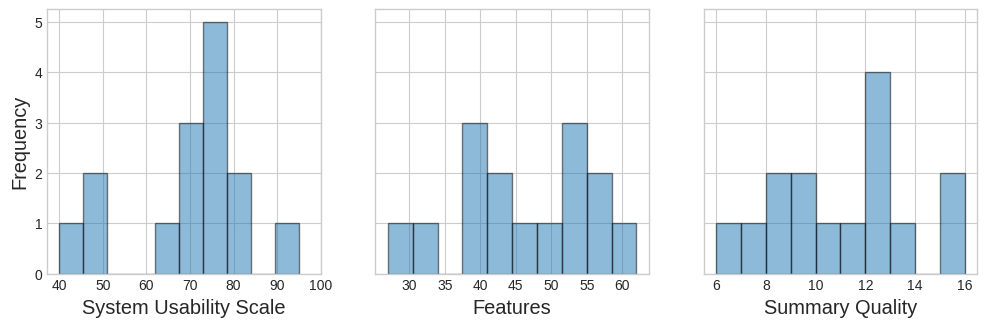}    
    \caption{Distribution of the questionnaire scores obtained during the user study. The possible range for each one of the scores is the following: \emph{System Usability Scale} ranges from 0 to 100, with a mean of 69.67 and a median of 75; the \emph{Features} score ranges from 0 to 64 with a mean of 45.87 and a median of 45; and the \emph{Summary Quality} ranges from 0 to 16 with a mean of 10.67 and a median of 11. These results show that the users perceived the system as useful and well-designed and that the summaries it produces are adequate for the task.}
    \label{fig:scores}
\end{figure}

The frequency tables of the responses for the \emph{System Usability Scale} questionnaire, the \emph{Features} questionnaire, and the Summary Quality questionnaire can be found in the code repository. To make it easier to understand the responses from the questionnaires, we computed a score for the Features and Summary Quality parts in the same fashion as for the System Usability Scale: the questions with positive wording have a value from 0 to 4, depending on their position on the scale. In contrast, the questions with negative wording have a value from 4 to 0, again depending on their position on the scale. The distribution of the scores obtained during the user study is shown in Fig.~\ref{fig:scores}. 

%%%%%%%%%%%%%%%%%%%%%%%%%%%%%%%%%%%%%%%%%%%%%%%%%%%%
\section{Discussion}
%%%%%%%%%%%%%%%%%%%%%%%%%%%%%%%%%%%%%%%%%%%%%%%%%%%%

%%%%%%%%%%%%%%%%%%%%%%%%%%%%%%%%%%%%%%%%%%%%%%%%%%%%
\subsection{Questionnaire Responses}
%%%%%%%%%%%%%%%%%%%%%%%%%%%%%%%%%%%%%%%%%%%%%%%%%%%%

Overall, \emph{QuOTeS} received a positive response across users, as the questionnaires show that the system seems to fulfill its purpose. Most of the time, the participants reported that the sentences recommended by the system seemed relevant and that the summaries appeared succinct, concise, and complete. Participants felt they understood the system's task and how it works. Furthermore, they felt that the components of the system were useful. Nonetheless, the system can be improved in the following ways:

\begin{itemize}
    \item As shown by the last question of the \emph{System Usability Scale} questionnaire, participants felt that they needed to learn many things before using the system. This is understandable, as \emph{QuOTeS} is based on several concepts which are very specific to Natural Language Processing and Information Retrieval: the task of Query-Focused Summarization itself, the concept of embedding documents as points in space, and the concept of training a Machine Learning classifier on the fly to adapt it to the needs of the user. Nonetheless, knowledge of these concepts is not strictly required to obtain useful insights from the system.

    \item As shown by the \emph{Features} questionnaire, the system can still be improved in terms of speed. Also, the users felt it was unclear what the different settings do and how to interpret the information in the plots. This may be improved with a better deployment and a better introductory tutorial that provides use cases for each one of the options in the settings: giving the user some guidance about when it is best to use word unigrams, character trigrams, and Sentence-BERT embeddings would facilitate picking the correct options.
\end{itemize}

\begin{figure}[ht]
    \centering
    \begin{subfigure}[m]{0.32\textwidth}
        \includegraphics[width=\textwidth]{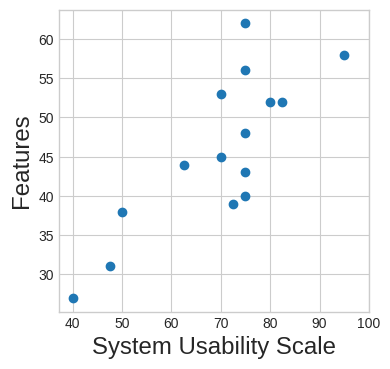}
    \end{subfigure}
    \hfill
    \begin{subfigure}[m]{0.32\textwidth}
        \includegraphics[width=\textwidth]{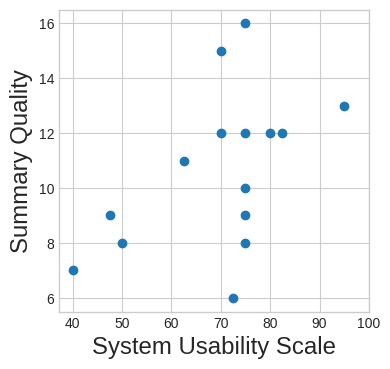}
    \end{subfigure}
    \hfill
    \begin{subfigure}[m]{0.32\textwidth}
        \includegraphics[width=\textwidth]{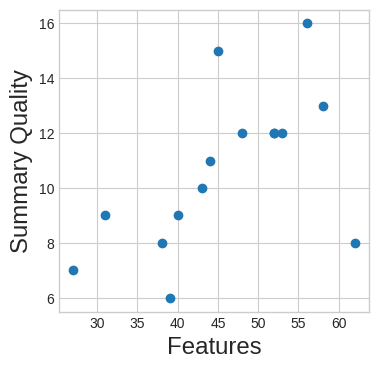}
    \end{subfigure}
    % \includegraphics{}
    % \caption{Relationship between the different scores computed from the responses of the user study. All the scores show a clear, positive relationship with each other with few outliers. This is to be expected since all of them are subjective and were measured immediately after each user tried the system. The most interesting observation is that there are two subgroups in the \emph{System Usability Scale}/\emph{Summary Quality} scatterplot: one in which the two scores both grow together and one in which the usability remains constant, and the summary quality varies wildly. This shows that the users found the system useful even when their query was unrelated to their document collection.}
    \caption{Relationship between the scores computed from the questionnaires.}
    \label{fig:relationship_scores}
\end{figure}

The relationship between the different scores computed from the responses of the user study is shown in Fig.~\ref{fig:relationship_scores}. All the scores show a clear, positive relationship with each other, with some outliers. The relationships found here are expected because all these scores are subjective and measure similar aspects of the system. Of all of them, the relationship between the System Usability Scale and the Summary Quality is the most interesting: it shows two subgroups, one in which the usability remains constant and the summary quality varies wildly, and another in which they both grow together. This may suggest that for some users, the query is so different from the collection that, although the system feels useful, they are dissatisfied with the results.

%%%%%%%%%%%%%%%%%%%%%%%%%%%%%%%%%%%%%%%%%%%%%%%%%%%%
\subsection{Analysis of the Labels Collected During the User Study}
%%%%%%%%%%%%%%%%%%%%%%%%%%%%%%%%%%%%%%%%%%%%%%%%%%%%

To further evaluate the performance of \emph{QuOTeS}, we estimated the Precision and Topic Concentration using the data labeled by the users. To compute the Precision, we divided the number of sentences labeled as relevant over the total number of sentences shown to the user. To compute the Topic Concentration, we followed the approach from \cite{topic_concentration}, using the Kullback-Leibler Divergence \cite{kl_divergence} between the unigram-based vocabulary of the document collection and the unigram-based vocabulary of the query-focused summary produced. 

\begin{figure}[ht]
    \centering
    \begin{subfigure}[m]{0.63\textwidth}
        \includegraphics[width = \textwidth]{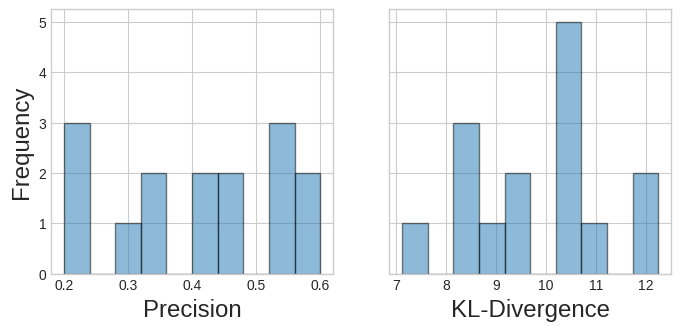}
        % \label{fig:dataset_metrics}
    \end{subfigure}
    \hfill
    \begin{subfigure}[m]{0.35\textwidth}
        \includegraphics[width = \textwidth]{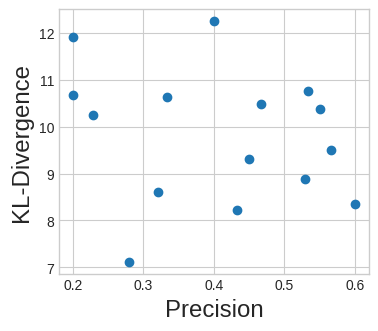}
    \end{subfigure}
    % \caption{Distributions of the Precision of the system (left) and the Kullback-Leibler Divergence between the word unigram distribution of the document collections and the summaries produced (center), along with their relationship (right). The plot shows that the system's recommendations are relevant around 40\% of the time and that the queries submitted by the users during the user study are not similar to the overall topic of the document collections. The relationship between these two metrics is unclear, implying that not all the examples with high Precision come from queries related to document collection. The relationship is also somewhat negative, which suggests that as the query becomes closer to the document collection, the recommendations from the system are more accurate.}
    \caption{Distributions of the Precision of the system (left) and the Kullback-Leibler Divergence between the word unigram distribution of the document collections and the summaries produced (center), along with their relationship (right).}
    \label{fig:dataset_metrics}
\end{figure}

The distributions of the Precision and KL-Divergence, along with their relationship, are shown in Fig.~\ref{fig:dataset_metrics}. The relationship between the two metrics is noisy, but it is somewhat negative, suggesting that as the KL-Divergence decreases, the Precision increases. This result makes sense because the KL-Divergence measures how much the query deviates from the contents of the document collection.

\begin{figure}[ht]
    \centering
    \begin{subfigure}[t]{0.45\textwidth}
        \includegraphics[width = \textwidth]{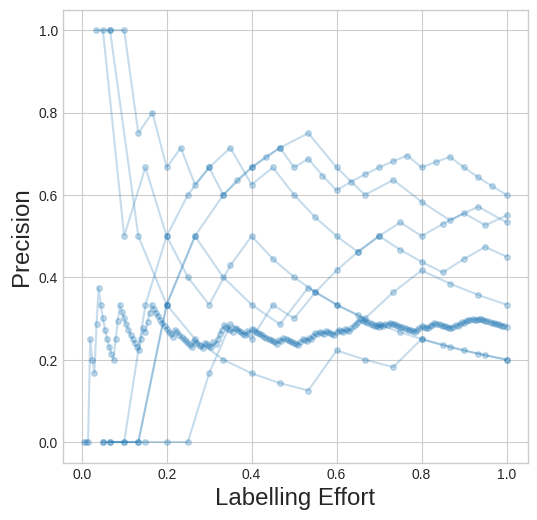}
        % \caption{Precision as a function of the Labeling Effort for each one of the participants in the user study.}
        % \label{fig:precision_effort}
    \end{subfigure}
    \hfill
    \begin{subfigure}[t]{0.45\textwidth}
        \includegraphics[width = \textwidth]{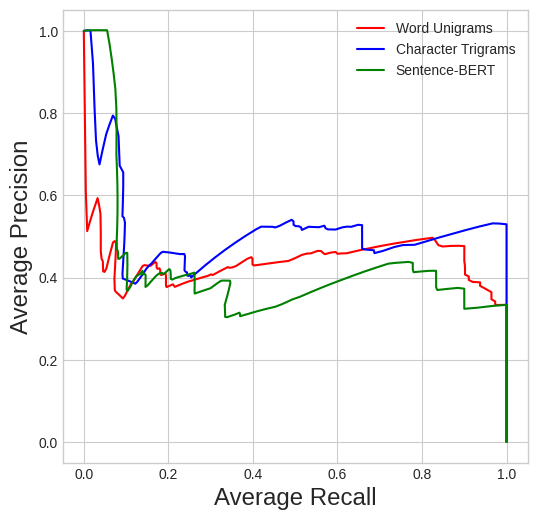}
        % \caption{Average Precision-Recall Curve of the different embeddings after removing the interactive component of \emph{QuOTeS}.}
        % \label{fig:precision_recall}

    \end{subfigure}
    \caption{Precision of the system. Precision as a function of the Labeling Effort for each one of the participants in the user study (left). Average Precision-Recall Curve of the different embeddings after removing the interactive component of \emph{QuOTeS} (right).}
    \label{fig:precision}
\end{figure}

On the other hand, Precision is displayed as a function of the Labeling Effort for each one of the participants in the user study in Fig.~\ref{fig:precision}. We computed the Labeling Effort as the fraction of sentences reviewed by the user. The system displays a stable average Precision of 0.39, which means that, on average, two out of five recommendations from the system are relevant. There appear to be two classes of users: in the first class, the system starts displaying a lot of relevant sentences, and the Precision drops as the system retrieves them; in the second class, the story is entirely the opposite: the system starts with very few correct recommendations, but it improves quickly as the user explores the collection.

% \begin{figure}[ht]
%     \centering
%     \includegraphics[width = 0.5\textwidth]{images/precision_effort.png}
%     % \caption{Precision as a function of the Labeling Effort for each one of the participants in the user study. The system displays a stable average Precision of 0.39, which means that, on average, two out of five recommendations from the system are relevant. There appear to be two classes of users: in the first class, the system starts displaying a lot of relevant sentences, and the Precision drops as the system retrieves the relevant sentences; while for the second class, the story is completely the opposite: the system starts with very few correct recommendations, but it improves quickly as the user explores the collection.}
%     \caption{Precision as a function of the Labeling Effort for each one of the participants in the user study.}
%     \label{fig:precision_effort}
% \end{figure}

\begin{figure}[ht]
    \centering
    \includegraphics[width = 0.9\textwidth]{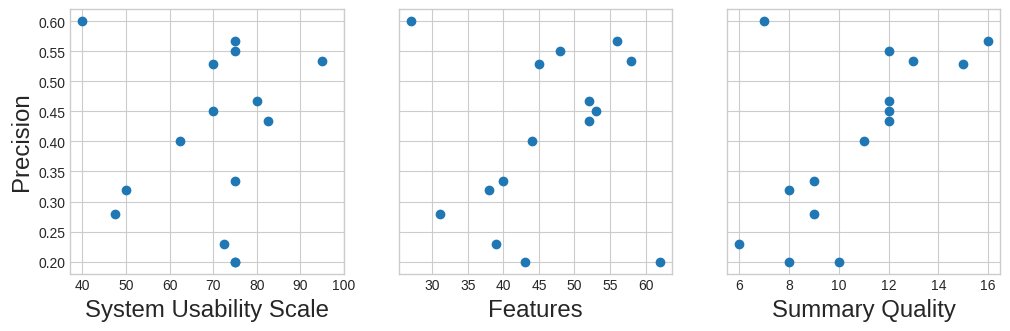}
    % \caption{Relationship between the Precision of the system and the questionnaire scores obtained from the user study. All the relationships are positive, clear, and have few outliers. This is expected since users probably mark the system highly after seeing it recommending relevant sentences. The outlier at the top left of the plots is very interesting, as it has the highest Precision of all the examples collected but received very low scores across all the questionnaires.}
    \caption{Relation between the Precision of the system and the questionnaire scores. }
    \label{fig:precision_vs}
\end{figure}

The relationships between the Precision and the scores obtained from the questionnaires in the user study are shown in Fig.~\ref{fig:precision_vs}. Precision is well correlated with all the other scores, which is expected since it is the first metric perceived by the user, even before answering the questionnaires. An outlier is very interesting: one of the users gave the system low scores in terms of the questionnaires, despite having the highest Precision of the dataset. The labels produced by this user display a lower Divergence than usual, which means that his query was much closer to the collection than most users, as shown in Fig.~\ref{fig:dataset_metrics}. This could mean that he/she could already have excellent previous knowledge about the document collection. Therefore, although the system was retrieving relevant sentences, it was not giving the user any new knowledge.

\begin{figure}[ht]
    \centering
    \includegraphics[width = 0.9\textwidth]{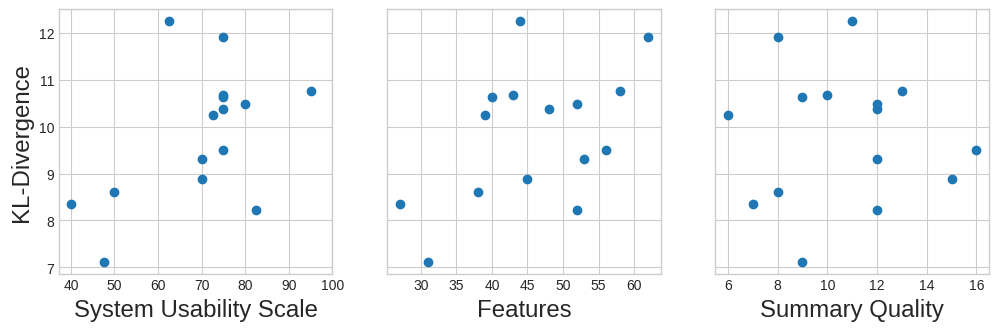}
    % \caption{Relationship between the Kullback-Leibler Divergence between the word unigram distribution of the document collection and produced summaries versus the questionnaire scores obtained in the user study. The relationship between the Divergence and the \emph{SUS} and the \emph{Features} scores are clear and positive, suggesting that as the query moves away from the document collection, the system is perceived as more useful and better designed. On the other hand, the relationship with the \emph{Summary Quality} score is noisy and somewhat neutral, suggesting that the perceived quality of the produced summaries is somewhat independent of the relatedness of the query and the document collection.}
    \caption{Relationship between the Kullback-Leibler Divergence between the word unigram distribution of the document collection and produced summaries versus the questionnaire scores obtained in the user study.}
    \label{fig:divergence_vs}
\end{figure}

The relationship between the Divergence and the scores is shown in Fig.~\ref{fig:divergence_vs}. The relationship shown is noisier than the ones involving Precision. Although the System Usability Scale and Features scores show a positive relationship with the Divergence, this is not the case with the Summary Quality. This suggests that to have a high-quality summary, it is necessary to start with a collection close to the query. Another interesting point is that these relationships suggest that the system is perceived as more useful and better designed as the query deviates from the document collection.

To finalize our evaluation of \emph{QuOTeS}, we measured its performance using the \emph{(Query, Document Collection)} pairs collected during the user study. As a baseline, we used the traditional \emph{Vector Space Model}, which is equivalent to disabling the \emph{Machine Learning Classifier} component of \emph{QuOTeS} (as shown in Fig.~\ref{fig:overview}). We evaluated the three variations of the baseline system as they appear inside \emph{QuOTeS}. The performance obtained by this baseline is shown in Fig.~\ref{fig:precision}.

Even when using Sentence-BERT embeddings, the performance of the baseline system is markedly inferior compared to that of \emph{QuOTeS}, as shown in Fig.~\ref{fig:precision}. Although the Sentence-BERT embeddings start with a much higher Precision than the traditional embeddings, they quickly deteriorate as the score threshold increases, while the traditional embeddings catch up in terms of Precision with the same level of Recall. However, since none of these models obtained a satisfactory performance, it is clear that using \emph{QuOTeS} enabled the users to find much more relevant sentences than they could have found otherwise. This highlights the importance of the Continuous Active Learning protocol in \emph{QuOTeS}, as it enables the system to leverage the feedback from the user, so the results do not depend entirely on the embeddings produced by the language model.

% \begin{figure}[ht]
%     \centering
%     \includegraphics[width = 0.5
%     \textwidth]{images/precision_recall.png}
%     % \caption{Average Precision-Average Recall Curve of the different embedding techniques after removing the interactive component of \emph{QuOTeS}. The Precision and Recall here are averaged over the labels of all the (Query, Document Collection) pairs collected during the user study. Although the Sentence-BERT embeddings start with a much higher Precision than the traditional embeddings, they deteriorate very quickly as the score threshold increases, while the traditional embeddings can catch up in terms of Precision with the same level of Recall. However, since none of these models obtained a satisfactory performance, it is clear that using \emph{QuOTeS} enabled the users to find much more relevant sentences than they could have found otherwise.}
%     \caption{Average Precision-Average Recall Curve of the different embedding techniques after removing the interactive component of \emph{QuOTeS}.}
%     \label{fig:precision_recall}
% \end{figure}

%%%%%%%%%%%%%%%%%%%%%%%%%%%%%%%%%%%%%%%%%%%%%%%%%%%%
\subsection{Limitations}
%%%%%%%%%%%%%%%%%%%%%%%%%%%%%%%%%%%%%%%%%%%%%%%%%%%%

Although our experimental results are promising, the system we propose has two main limitations, given the complexity of the task and the amount of resources needed to produce benchmarks for this topic:

\begin{itemize}
    \item First, the purpose of \emph{QuOTeS} is not to provide fully automatic summaries since it is hard to guarantee that all the relevant sentences were retrieved in the process. Instead, its purpose is to point users in the right direction so that they can find the relevant information in the original documents.

    \item And second, the summaries produced by the system can still be improved using traditional techniques from Automatic Summarization. For example, their sentences in the summary could be reordered or removed to improve fluency and conciseness. These aspects would be beneficial if the goal is to produce a fully-automatic summary of the collection of articles.   
\end{itemize}

%%%%%%%%%%%%%%%%%%%%%%%%%%%%%%%%%%%%%%%%%%%%%%%%%%%%
\section{Conclusions and Future Work}
%%%%%%%%%%%%%%%%%%%%%%%%%%%%%%%%%%%%%%%%%%%%%%%%%%%%

In this paper, we introduce \emph{QuOTeS}, a system for Query-Focused Summarization of Scientific Documents designed to retrieve sentences relevant to a short paragraph, which takes the role of the query. \emph{QuOTeS} is an interactive system based on the Continuous Active Learning protocol that incorporates the user's feedback in the retrieval process to adapt itself to the user's query. 

After a comprehensive analysis of the questionnaires and labeled data obtained through a user study, we found that \emph{QuOTeS} provides a positive user experience and fulfills its purpose. Also, the experimental results show that including both the user's information need and feedback in the retrieval process leads to better results that cannot be obtained with the current non-interactive methods. 

For future work, we would like to conduct a more comprehensive user study where users read the whole papers and label the sentences manually, after which they could use \emph{QuOTeS} and compare the summaries produced. Another interesting future direction would be to compare the system heads-on with the main non-interactive methods from the literature on a large, standardized dataset.

%%%%%%%%%%%%%%%%%%%%%%%%%%%%%%%%%%%%%%%%%%%%%%%%%%%%
\section{Acknowledgements}
%%%%%%%%%%%%%%%%%%%%%%%%%%%%%%%%%%%%%%%%%%%%%%%%%%%%

% HIDDEN IN THIS DRAFT TO ENABLE BLIND REVIEW.

We thank the Digital Research Alliance of Canada (\url{https://alliancecan.ca/en}), CIUNSa (Project B 2825), CONICET (PUE 22920160100056CO, PIBAA 2872021010 1236CO), MinCyT (PICT PRH-2017-0007), UNS (PGI 24/N051) and the Natural Sciences and Engineering Research Council of Canada (NSERC) for the resources provided to enable this research.

\bibliographystyle{unsrt}
% \bibliography{main}
\bibliography{references}

\end{document}